\documentclass[useAMS,usenatbib]{mn2e}
\usepackage{graphicx}
\usepackage{times}
\usepackage{amssymb}
\usepackage{natbib}
\def\gsim { \lower .75ex \hbox{$\sim$} \llap{\raise .27ex \hbox{$>$}} }
\def\lsim { \lower .75ex \hbox{$\sim$} \llap{\raise .27ex \hbox{$<$}} }

\begin{document}

\title[Satellites on Extreme Orbits] 
{Cosmic M\'enage \`a Trois: The Origin of Satellite Galaxies \\ on Extreme Orbits}

\author[Sales, Navarro, Abadi \& Steinmetz]{Laura V. Sales$^{1,2}$, Julio F. Navarro,$^{3,4}$\thanks{Fellow of the Canadian Institute for Advanced Research.} Mario G. Abadi $^{1,2,3}$ 
and Matthias Steinmetz$^{5}$
\\
$^{1}$ Observatorio Astron\'omico, Universidad Nacional de C\'ordoba, Laprida
854, 5000 C\'ordoba, Argentina.
\\
$^{2}$ Instituto de Astronom\'{\i}a Te\'orica y Experimental, Conicet, Argentina.
\\
$^{3}$Department of Physics and Astronomy, University of Victoria, Victoria, BC V8P 5C2,
Canada\\
$^{4}$Max-Planck Institut f\"ur Astrophysik, Karl-Schwarzschild Strasse 1,
Garching, D-85741, Germany\\
$^{5}$Astrophysikalisches Institut Potsdam, An der Sternwarte 16, Potsdam 14482, Germany\\
}

\maketitle

\begin{abstract}
We examine the orbits of satellite galaxies identified in a suite of
N-body/gasdynamical simulations of the formation of $L_*$ galaxies in
a $\Lambda$CDM universe. The numerical resolution of the simulations
allows us to track in detail the orbits of the $\sim $ ten brightest
satellites around each primary. Most satellites follow conventional
orbits; after turning around, they accrete into their host halo and
settle on orbits whose apocentric radii are steadily eroded by
dynamical friction. As a result, satellites associated with the
primary are typically found within its virial radius, $r_{\rm vir}$,
and have velocities consistent with a Gaussian distribution with mild
radial anisotropy. However, a number of outliers are also present. We
find that a surprising number (about one-third) of satellites
identified at $z=0$ are on unorthodox orbits, with apocenters that
exceed their turnaround radii. These include a number of objects with
extreme velocities and apocentric radii at times exceeding $\sim 3.5\,
r_{\rm vir}$ (or, e.g., $\gsim \, 1$ Mpc when scaled to the Milky
Way). This population of satellites on extreme orbits consists
typically of the faint member of a satellite pair whose kinship is
severed by the tidal field of the primary during first approach. Under
the right circumstances, the heavier member of the pair remains bound
to the primary, whilst the lighter companion is ejected onto a
highly-energetic orbit. Since the concurrent accretion of multiple
satellite systems is a defining feature of hierarchical models of
galaxy formation, a fairly robust prediction of this scenario is that
at least some of these extreme objects should be present in the Local
Group. We speculate that this three-body ejection mechanism may be the
origin of (i) some of the newly discovered high-speed satellites
around M31 (such as Andromeda XIV); (ii) some of the distant
fast-receding Local Group members, such as Leo I; and (iii) the oddly
isolated dwarf spheroidals Cetus and Tucana in the outskirts of the
Local Group. Our results suggest that care must be exercised when
using the orbits of the most weakly bound satellites to place
constraints on the total mass of the Local Group.
\end{abstract}


\section{Introduction}
\label{sec:intro}

The study of Local Group satellite galaxies has been revolutionized by
digital imaging surveys of large areas of the sky. More than a dozen
new satellites have been discovered in the past couple of years
\citep{zucker04,zucker06,willman05b,martin06,belokurov06,
belokurov07,irwin07,majewski07},
 due in large part to the completion of the Sloan Digital Sky
Survey \citep{york00,strauss02} and to concerted campaigns designed to image in detail
the Andromeda galaxy and its immediate surroundings (\citealt{ibata01,ferguson02,
reitzel02,mcconnachie03,rich04,guhathakurta06,gilbert06,chapman06},
Ibata et al. 2007 submitted). The
newly discovered satellites have extended the faint-end of the galaxy
luminosity function down to roughly $\sim 10^3 \,
L_{\odot}$, and are likely to provide important constraints regarding
the mechanisms responsible for ``lighting up'' the baryons in low-mass
halos. These, in turn, will serve to validate (or falsify) the various
theoretical models attempting to reconcile the wealth of
``substructure'' predicted in cold dark matter (CDM) halos with the
scarcity of luminous satellites in the Local Group 
\citep[see, e.g.][]{klypin99b,bullock00, benson02,stoehr02,
kazantzidis04,kravtsov04,penarrubia07}.

At the same time, once velocities and distances are secured for the
newly-discovered satellites, dynamical studies of the total mass and
spatial extent of the Local Group will gain new impetus. These studies
have a long history \citep{littleandtremaine87,zaritsky89,kochanek96,wilkinson99,
evans00,battaglia05}, but their results have traditionally
been regarded as tentative rather than conclusive, particularly
because of the small number of objects involved, as well as the
sensitivity of the results to the inclusion (or omission) of one or
two objects with large velocities and/or distances \citep{zaritsky89,
kochanek96,sakamoto03}. An enlarged satellite sample will likely 
make the conclusions of satellite dynamical studies more 
compelling and robust.

To this end, most theoretical work typically assumes that satellites are in
equilibrium, and use crafty techniques to overcome the limitations of
small-N statistics when applying Jeans' equations to estimate masses
\citep[see, e.g.,][]{littleandtremaine87,wilkinson99,evans00}. With increased
sample size, however, follow enhanced
opportunities to discover satellites on unlikely orbits; i.e.,
dynamical ``outliers'' that may challenge the expectations of
simple-minded models of satellite formation and evolution.  It is
important to clarify the origin of such systems, given their
disproportionate weight in mass estimates.

One issue to consider is that the assumption of equilibrium must break
down when considering outliers in phase space. This is because the
finite age of the Universe places an upper limit to the orbital period
of satellites observed in the Local Group; high-speed satellites have
typically large apocenters and long orbital periods, implying that
they cannot be dynamically well-mixed and casting doubts on the
applicability of Jeans' theorem-inspired analysis tools.

To make progress, one possibility is to explore variants of the
standard secondary infall model \citep{gunnandgott72, gott75, 
gunn77,fillmore84}, where satellites are assumed to recede initially with
the universal expansion, before turning around and collapsing onto the
primary due to its gravitational pull. This is the approach adopted by
\citet{zaritskyandwhite94}  in order to interpret statistically the
kinematics of observed satellite samples without assuming well-mixed
orbits and taking into account the proper timing and phase of the
accretion process.

\begin{center}
\begin{figure*}
\includegraphics[width=100mm]{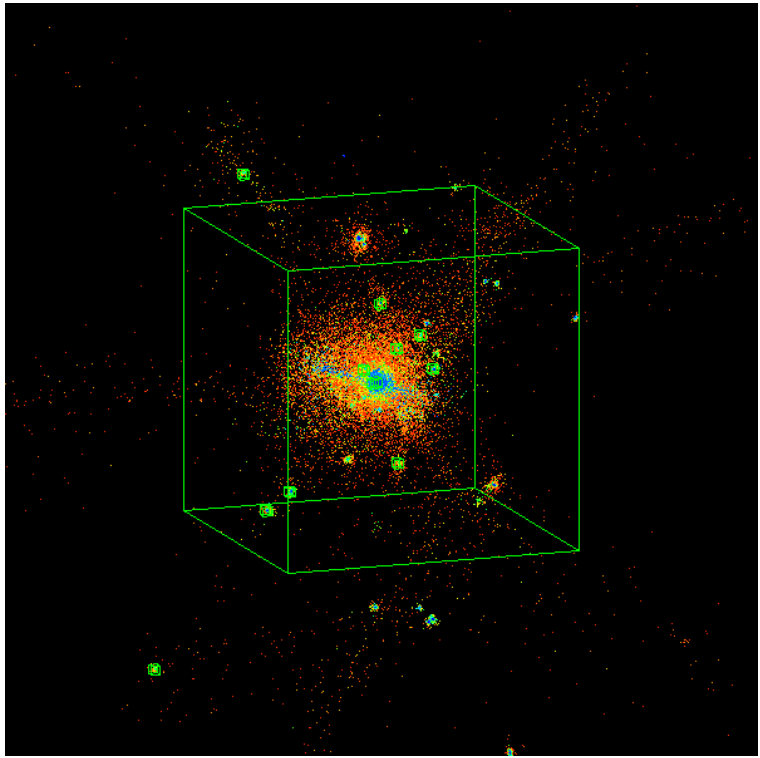}
\caption{Star particles in one of our simulations, shown at
$z=0$. Particles are colored according to the age of the star; blue
means a star is younger than $\simeq 1$ Gyr, red that it is older than
$\simeq 10$ Gyr. The large box is $2\, r_{\rm vir}$ ($632$ kpc) on a
side and centered on the primary galaxy. More than $85\%$ of all stars
are in the inner regions of the primary, within about $\sim 20$ 
kpc from the center \citep[for more details see][]{abadi06}.
surround the satellites ``associated'' with the primary galaxy; i.e.,
satellites that have been within $r_{\rm vir}$ in the past. Note that
a few ``associated'' satellites lie well beyond the virial boundary of
the system. Two of these satellites are highlighted for analysis in
Figures~\ref{fig:orbesc} and \ref{fig:orbxyesc}.}
\label{fig:xyzsat}
\end{figure*}
\end{center}

In the secondary infall accretion sequence, satellites initially
farther away accrete later, after turning around from larger
turnaround radii. The turn-around radius grows with time, at a rate
the depends on the mass of the primary and its environment, as well as
on the cosmological model.  Three distinct regions surround a system
formed by spherical secondary infall \citep[see, e.g.,][]{bertschinger85,
navarroandwhite93}: (i) an outer region beyond the current
turnaround radius where satellites are still expanding away, (ii) an
intermediate region containing satellites that are approaching the
primary for the first time, and (iii) an inner, ``virialized'' region
containing all satellites that have turned around at earlier times and
are still orbiting around the primary.  To good approximation, the
latter region is delineated roughly by the conventional virial radius
of a system{\footnote{We define the {\it virial} radius, $r_{\rm
vir}$, of a system as the radius of a sphere of mean density $\simeq
\Delta_{\rm vir}(z)$ times the critical density for closure. This
definition defines implicitly the virial mass, $M_{\rm vir}$, as that
enclosed within $r_{\rm vir}$, and the virial velocity, $V_{\rm vir}$,
as the circular velocity measured at $r_{\rm vir}$. We compute
$\Delta_{\rm vir}(z)$ using $\Delta_{vir}(z)=18\pi^2+82f(z)-39f(z)^2$, where
$f(z)=[\Omega_0(1+z)^3/(\Omega_0(1+z)^3+\Omega_\Lambda))]-1$ and
$\Omega_0=\Omega_{\rm CDM}+\Omega_{\rm
bar}$ \citep{bryanandnorman98}, and is $\sim 100$ at $z=0$}}, $r_{\rm vir}$;
the turnaround radius is of order $r_{\rm ta} \sim 3 \,r_{\rm vir}$
\citep[see, e.g.][]{white93}.

We note a few consequences of this model. (a) Satellites outside the
virial radius are on their first approach to the system and thus have
not yet been inside $r_{\rm vir}$. (b) Satellites inside the virial
radius have apocentric radii that typically do not exceed $r_{\rm
vir}$. (c) The farther the turnaround radius the longer it takes for a
satellite to turn around and accrete and the higher its orbital
energy. (d) Satellites with extreme velocities will, in general, be
those completing their first orbit around the primary. Velocities will
be maximal near the center, where satellites may reach speeds as high
as $\sim 3 \,V_{\rm vir}$. (e) Since all satellites associated with
the primary are bound (otherwise they would not have turned around and
collapsed under the gravitational pull of the primary), the velocity
of the highest-speed satellite may be used to estimate a lower limit
to the escape velocity at its location and, thus, a lower bound to the
total mass of the system.

Hierarchical galaxy formation models, such as the current $\Lambda$CDM
paradigm, suggest further complexity in this picture. Firstly,
although numerical simulations show that the sequence of expansion,
turnaround and accretion of satellites described above is more or less
preserved in hierarchical models, the evolution is far from
spherically symmetric
\citep{navarro94,ghigna98,jing02,bailin05,knebe06}.  Much of the mass
(as well as many of the satellites) is accreted through filaments of
matter embedded within sheets of matter formation \citep[see,
e.g.,][]{navarro04}. The anisotropic collapse pattern onto a primary
implies that the turnaround ``surface'' won't be spherical and that
the virial radius may not contain {\it all} satellites that have
completed at least one orbit around the primary \citep[see,
e.g.,][]{balogh00,diemand07}.

More importantly for the purposes of this paper, in hierarchical
models galaxy systems are assembled by collecting smaller systems
which themselves, in turn, were assembled out of smaller units. This
implies that satellites will in general not be accreted in isolation,
but frequently as part of larger structures containing multiple
systems. This allows for complex many-body interactions to take place
during approach to the primary that may result in substantial
modification to the orbits of accreted satellites.

We address this issue in this contribution using N-body/gasdynamical
simulations of galaxy formation in the current $\Lambda$CDM paradigm.
We introduce briefly the simulations in \S~\ref{sec:numexp}, and
analyze and discuss them in \S~\ref{sec:analysis}. We speculate on
possible applications to the Local Group satellite population in
\S\ref{sec:LG} and conclude with a brief summary in \S~\ref{sec:conc}.

\section{The Numerical Simulations}
\label{sec:numexp}

We identify satellite galaxies in a suite of eight simulations of the
formation of $L_*$ galaxies in the $\Lambda$CDM scenario. This series
has been presented by Abadi, Navarro \& Steinmetz (2006), and follow
the same numerical scheme originally introduced by 
\citet{steinmetzandnavarro02}. The ``primary'' galaxies in 
these simulations have been analyzed in detail in several recent 
papers, which the interested reader may wish to consult for 
details \citep{abadi03a,abadi03b,meza03,meza05,navarro04}.
We give a brief outline below for completeness.

Each simulation follows the evolution of a small region of the
universe chosen so as to encompass the mass of an $L_{*}$ galaxy
system. This region is chosen from a large periodic box and
resimulated at higher resolution preserving the tidal fields from the
whole box. The simulation includes the gravitational effects of dark
matter, gas and stars, and follows the hydrodynamical evolution of the
gaseous component using the Smooth Particle Hydrodynamics (SPH)
technique \citep{steinmetz96}. We adopt the following cosmological
parameters for the $\Lambda$CDM scenario: $H_0=65$ km/s/Mpc,
$\sigma_8=0.9$, $\Omega_{\Lambda}=0.7$, $\Omega_{\rm CDM}=0.255$,
$\Omega_{\rm b}=0.045$, with no tilt in the primordial power
spectrum. 

All re-simulations start at redshift $z_{\rm init}=50$, have force
resolution of order $1$ kpc, and the mass resolution is chosen so that
each galaxy is represented on average, at $z=0$, with $\sim 50,000$
dark matter/gas particles. Gas is turned into stars at rates
consistent with the empirical Schmidt-like law of \citet{kennicutt98}.
Because of this, star formation proceeds efficiently only in
high-density regions at the center of dark halos, and the stellar
components of primary and satellite galaxies are strongly segregated
spatially from the dark matter.

Each re-simulation follows a single $\sim L_*$ galaxy in detail, and
resolves as well a number of smaller, self-bound systems of stars,
gas, and dark matter we shall call generically ``satellites''. We
shall hereafter refer to the main galaxy indistinctly as ``primary''
or ``host''.  The resolved satellites span a range of luminosities,
down to about six or seven magnitudes fainter than the primary. Each
primary has on average $\sim 10$ satellites within the virial radius.

Figure~\ref{fig:xyzsat} illustrates the $z=0$ spatial configuration of
star particles in one of the simulations of our series. Only star
particles are shown here, and are colored according to their age:
stars younger than $\simeq 1$ Gyr are shown in blue; those older than
$\simeq 10$ Gyr in red. The large box is centered on the primary and
is $2\, r_{\rm vir}$ ($632$ kpc) on a side. The ``primary'' is
situated at the center of the large box and contains most of the
stars. Indeed, although not immediately apparent in this rendition,
more than $85\%$ of all stars are within $\sim 20$ kpc from the
center. Outside that radius most of the stars are old and belong to
the stellar halo, except for a plume of younger stars stripped from a
satellite that has recently merged with the primary. Satellites
``associated'' with the primary (see \S~\ref{ssec:convorb} for a
definition) are indicated with small boxes. Note that a few of them
lie well beyond the virial radius of the primary.

A preliminary analysis of the properties of the simulated satellite
population and its relation to the stellar halo and the primary galaxy
has been presented in Abadi, Navarro \& Steinmetz (2006) and Sales et
al (2007, submitted), where the interested reader may find further details.
\begin{center}
\begin{figure}
\includegraphics[width=84mm]{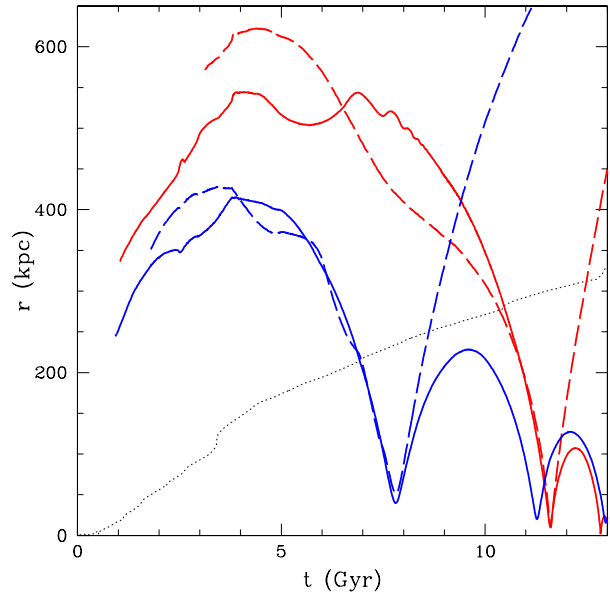}
\caption{ Distance to the primary as a function of
time for four satellites selected in one of our simulations. The four
satellites are accreted into the primary in two pairs of unequal
mass. The heavier satellite of the pair, shown by solid lines, follows
a ``conventional'' orbit: after turning around from the universal
expansion, it accretes into the primary on a fairly eccentric orbit
which becomes progressively more bound by the effects of dynamical
friction. Note that, once accreted, these satellites on
``conventional'' orbits do not leave the virial radius of the primary,
which is shown by a dotted line. The light member of the pair, on
the other hand, is ejected from the system as a result of a three-body
interaction between the pair and the primary during first
approach. One of the ejected satellites shown here is the ``escaping''
satellite identified in Figure~\ref{fig:rvMW}; the other is the most
distant ``associated'' satellite in that Figure. The latter is still
moving toward apocenter at $z=0$, which we estimate to be as far as
$\sim 3.5\, r_{\rm vir}$.}
\label{fig:orbesc}
\end{figure}
\end{center}

\section{Results and Discussion}
\label{sec:analysis}

\subsection{Satellites on conventional orbits}
\label{ssec:convorb}

The evolution of satellites in our simulations follows roughly the
various stages anticipated by our discussion of the secondary infall
model; after initially receding with the universal expansion, satellites turn around and are
accreted into the primary. Satellites massive enough to be well
resolved in our simulations form stars actively before accretion and,
by the time they cross the virial radius of the primary, much of their
baryonic component is in a tightly bound collection of stars at the
center of their own dark matter halos. 

The stellar component of a satellite is thus quite resilient to the
effect of tides and can survive as a self-bound entity for several
orbits. This is illustrated by the {\it solid lines} in
Figure~\ref{fig:orbesc}, which show, for one of our simulations, the
evolution of the distance to the primary of two satellites that turn
around and are accreted into the primary at different times. As
expected from the secondary infall model, satellites that are
initially farther away turn around later; do so from larger radii; and
are on more energetic orbits. After accretion (defined as the time
when a satellite crosses the virial radius of the primary), their
orbital energy and eccentricity are eroded by dynamical friction, and
these two satellites do not leave the virial radius of the primary,
shown by the dotted line in Figure~\ref{fig:orbesc}. Depending on
their mass and orbital parameters, some of these satellites merge with
the primary shortly after accretion, while others survive as
self-bound entities until $z=0$. For short, we shall refer to
satellites that, by $z=0$, have crossed the virial radius boundary at
least once as satellites ``associated'' with the primary.

The ensemble of surviving satellites at $z=0$ have kinematics
consistent with the evolution described above. This is illustrated in
Figure~\ref{fig:rvMW}, where we show the radial velocities of all
satellites as a function of their distance to the primary, scaled to
virial units. Note that the majority of ``associated'' satellites
(shown as circles in this figure) are confined within $r_{\rm vir}$,
and that their velocity distribution is reasonably symmetric and
consistent with a Gaussian (Sales et al 2007). The most recently
accreted satellites tend to have higher-than-average speed at all
radii, as shown by the ``crossed'' circles, which identify all
satellites accreted within the last $3$ Gyr.

Crosses (without circles) in this figure correspond to satellites that
have not yet been accreted into the primary.  These show a clear infall
pattern outside $r_{\rm vir}$, where the mean infall velocity
decreases with radius and approaches zero at the current turnaround
radius, located at about $3 \, r_{\rm vir}$. All of these properties
agree well with the expectations of the secondary infall model
discussed above.

\subsection{Three-body interactions and satellites on unorthodox orbits}

Closer examination, however, shows a few surprises. To begin with, a
number of ``associated'' satellites are found outside $r_{\rm
vir}$. As reported in previous work \citep[see, e.g.,][]{balogh00,moore04,
gill05,diemand07}, these are a minority ($\sim 15\%$ in
our simulation series), and have been traditionally linked to
departures from spherical symmetry during the accretion
process. Indeed, anisotropies in the mass distribution during
expansion and recollapse may endow some objects with a slight excess
acceleration or, at times, may push satellites onto rather tangential
orbits that ``miss'' the inner regions of the primary, where
satellites are typically decelerated into orbits confined within the
virial radius.

These effects may account for some of the associated satellites found
outside $r_{\rm vir}$ at $z=0$, but cannot explain why $\sim 33\%$ of
all associated satellites are today on orbits whose apocenters exceed
their turnaround radius. This is illustrated in
Figure~\ref{fig:rhist}, where we show a histogram of the ratio between
apocentric radius (measured at $z=0$; $r_{\rm apo}$) and turnaround
radius ($r_{\rm ta}$). The histogram highlights the presence of two
distinct populations: satellites on ``conventional'' orbits with
$r_{\rm apo}/r_{\rm ta} <1$, and satellites on orbital paths that lead
them well beyond their original turnaround radius.

Intriguingly, a small but significant fraction ($\sim 6\%$) of
satellites have extremely large apocentric radius, exceeding their
turnaround radius by $50\%$ or more. These systems have clearly been
affected by some mechanism that propelled them onto orbits
substantially more energetic than the ones they had followed until
turnaround. This mechanism seems to operate preferentially on low-mass
satellites, as shown by the dashed histogram in
Figure~\ref{fig:rhist}, which corresponds to satellites with stellar
masses less than $\sim 3\%$ that of the primary.

\begin{figure}
\includegraphics[width=84mm]{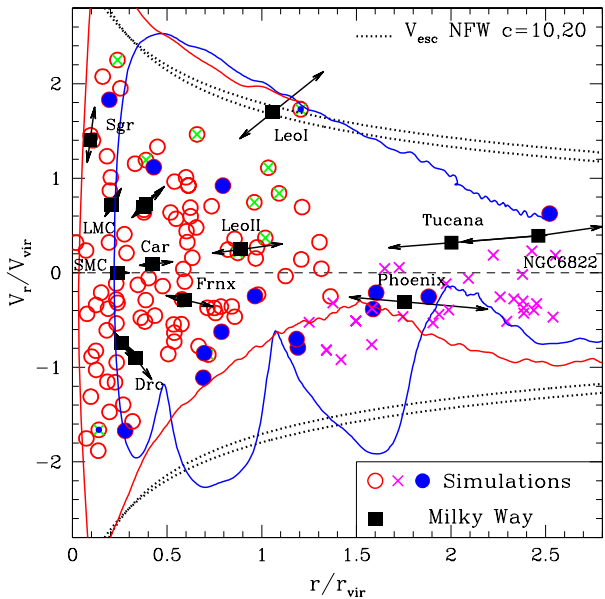}
\caption{Radial velocity of satellites versus distance to the
primary. Velocities are scaled to the virial velocity of the system,
distances to the virial radius. Circles denote ``associated''
satellites; i.e., those that have been {\it inside} the virial radius
of the primary at some earlier time. Crosses indicate satellites that
are on their first approach, and have never been inside $r_{\rm
vir}$. Filled circles indicate associated satellites whose apocentric
radii exceed their turnaround radius by at least $25\%$, indicating
that their orbital energies have been substantially altered during
their evolution. ``Crossed'' circles correspond to associated
satellites that have entered $r_{\rm vir}$ during the last $3$
Gyrs. The curves delineating the top and bottom boundaries of the
distribution show the escape velocity of an NFW halo with
concentration $c=10$ and $c=20$, respectively. 
Note that there is one satellite ``escaping'' the system with positive
radial velocity. Solid lines show the trajectories in the $r-V_r$ plane
of the two ''ejected'' satellites shown in figure \ref{fig:orbesc}.
Filled squares correspond to the fourteen brightest Milky
Way satellites, taken from \citet{vandenbergh99} (complemented with NED data for
the Phoenix, Tucana and NGC6822), and plotted assuming that $V_{\rm vir}^{\rm
MW} \sim 109$ km/s and $r_{\rm vir}^{\rm MW}=237$ kpc (see Sales et al
2007). Arrows indicate how the positions of MW satellites in this plot
would be altered if our estimate of $V_{\rm vir}^{\rm MW}$ (and,
consequently, $r_{\rm vir}^{\rm MW}$) is allowed to vary by $\pm 20\%$.}
\label{fig:rvMW}
\end{figure}

We highlight some of these objects in Figure~\ref{fig:rvMW}, using
``filled'' circles to denote ``associated'' satellites whose
apocenters at $z=0$ exceed their turnaround radii by at least
$25\%$. Two such objects are worth noting in this figure: one of them
is the farthest ``associated'' satellite, found at more than $\sim 2.5
\, r_{\rm vir}$ from the primary; the second is an outward-moving
satellite just outside the virial radius but with radial velocity
approaching $\sim 2\, V_{\rm vir}$. The latter, in particular, is an
extraordinary object, since its radial velocity alone exceeds the
nominal escape velocity{\footnote{The notion of binding energy and
escape velocity is ill-defined in cosmology; note, for example, that
the {\it whole universe} may be considered formally bound to any
positive overdensity in an otherwise unperturbed Eistein-de Sitter
universe. We use here the nominal escape velocity of an NFW model
\citep{nfw96,nfw97} to guide the
interpretation. This profile fits reasonably well the mass
distribution of the primaries inside the virial radius, and has a
finite escape velocity despite its infinite mass. Certainly satellites
with velocities exceeding the NFW escape velocity are likely to move
far enough from the primary to be considered true {\it escapers}.}} at
that radius. This satellite is on a trajectory which, for all
practical purposes, will remove it from the vicinity of the primary
and leave it wandering through intergalactic space.

The origin of these unusual objects becomes clear when inspecting
Figure~\ref{fig:orbesc}. The two satellites in question are shown with
{\it dashed lines} in this figure; each is a member of a bound {\it
pair} of satellites (the other member of the pair is shown with solid
lines of the same color). During first pericentric approach, the pair
is disrupted by the tidal field of the primary and, while one member
of the pair remains bound and follows the kind of ``conventional''
orbit described in \S~\ref{ssec:convorb}, the other one is ejected
from the system on an extreme orbit. The trajectories of these two
``ejected'' satellites in the $r$-$V_r$ plane are shown by the wiggly
lines in Figure~\ref{fig:rvMW}.

These three-body interactions typically involve the first pericentric
approach of a bound pair of accreted satellites and tend to eject the
lighter member of the pair: in the example of Figure~\ref{fig:orbesc},
the ``ejected'' member makes up, respectively, only $3\%$ and $6\%$ of
the total mass of the pair at the time of accretion. Other interaction
configurations leading to ejection are possible, such as an unrelated
satellite that approaches the system during the late stages of a
merger event, but they are rare, at least in our simulation series. We
emphasize that not all satellites that have gained energy during
accretion leave the system; most are just put on orbits of unusually
large apocenter but remain bound to the primary. This is shown by the
filled circles in Figure~\ref{fig:rvMW}; many affected satellites are
today completing their second or, for some, third orbit around the
primary.

\begin{center}
\begin{figure}
\includegraphics[width=84mm]{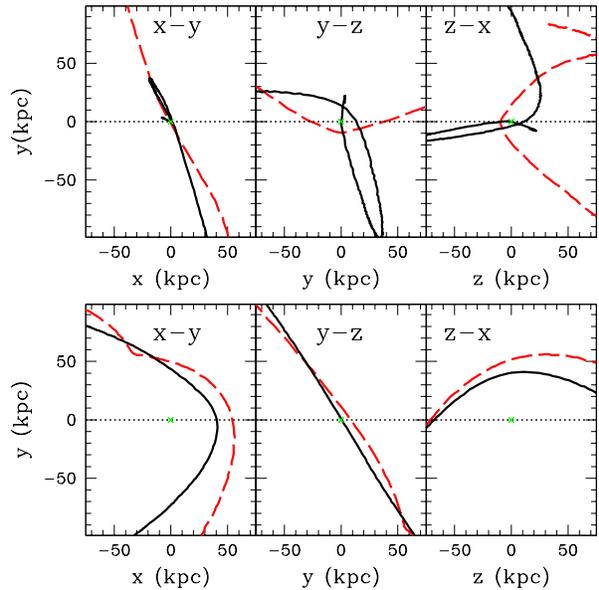}
\caption{Orbital paths for both pair of satellites shown in
Figure~\ref{fig:orbesc}. Upper (bottom) panel shows the pair that
accretes later (earlier) in that figure and shows the orbits in the
rest frame
of the primary. The coordinate system is chosen so that the angular
momentum of the primary is aligned with the $z$ axis. A solid curve
tracks the path of the heavier satellite; a dashed line follows the
satellite that is propelled into a highly energetic orbit after.}
\label{fig:orbxyesc}
\end{figure}
\end{center}

The ejection mechanism is perhaps best appreciated by inspecting the
orbital paths of the satellite pairs. These are shown in
Figure~\ref{fig:orbxyesc}, where the top (bottom) panels correspond to
the satellite pair accreted later (earlier) into the primary in
Figure~\ref{fig:orbesc}. Note that in both cases, as the pair
approaches pericenter, the lighter member (dashed lines) is also in
the process of approaching the pericenter of its own orbit around the
heavier member of the pair. This coincidence in orbital phase
combines the gravitational attraction of the two more massive members
of the trio of galaxies, leading to a substantial gain in orbital
energy by the lightest satellite, effectively ejecting it from the
system on an approximately radial orbit. The heavier member of the
infalling pair, on the other hand, decays onto a much more tightly
bound orbit.

Figure~\ref{fig:orbxyesc} also illustrates the complexity of orbital
configurations that are possible during these three-body
interactions. Although the pair depicted in the top panels approaches
the primary as a cohesive unit, at pericenter each satellite circles
about the primary in opposite directions: in the $y$-$z$ projection
the heavier member circles the primary {\it clockwise} whereas the
ejected companion goes about it {\it counterclockwise}. After
pericenter, not only do the orbits of each satellite have different
period and energy, but they differ even in the {\it sign} of their
orbital angular momentum. In this case it would clearly be very
difficult to link the two satellites to a previously bound pair on the
basis of observations of their orbits after pericenter.

Although not all ejections are as complex as the one illustrated in
the top panels of Figure~\ref{fig:orbxyesc}, it should be clear from
this figure that reconstructing the orbits of satellites that have
been through pericenter is extremely difficult, both for satellites
that are ejected as well as for those that remain bound. For example,
the massive member of the late-accreting pair in
Figure~\ref{fig:orbesc} sees its apocenter reduced by more than a
factor of $\sim 5$ from its turnaround value in a single pericentric
passage. Such dramatic variations in orbital energy are difficult to
reproduce with simple analytic treatments inspired on Chandrasekhar's
dynamical friction formula (Pe\~narrubia 2007, private communication).

\begin{center}
\begin{figure}
\includegraphics[width=84mm]{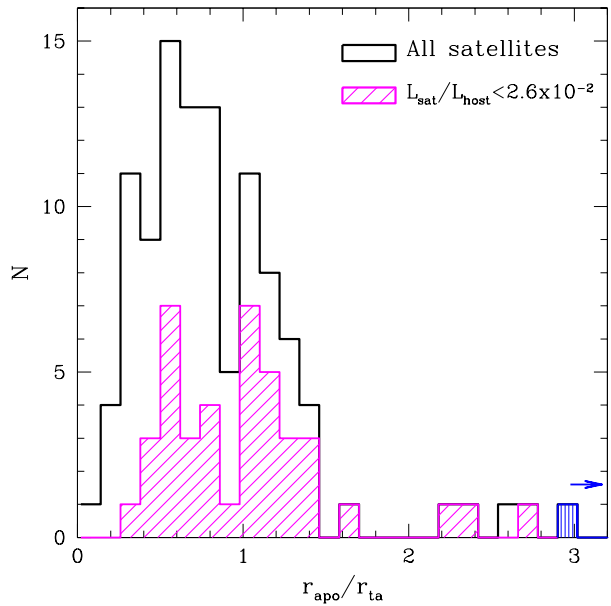}
\caption{Distribution of the ratio between the apocentric radius of
satellites (measured at $z=0$) and their turnaround radius, defined as
the maximum distance to the primary before accretion. Note the
presence of two groups. Satellites on ``conventional'' orbits have
$r_{\rm apo}/r_{\rm ta}<1$, the rest have been catapulted into
high-energy orbits by three-body interactions during first
approach. The satellite marked with a rightward arrow is the
``escaping'' satellite identified by a dot-centered circle in
Figure~\ref{fig:rvMW}; this system has nominally infinite $r_{\rm
apo}$. The dashed histogram highlights the population of low-mass
satellites; i.e., those with stellar masses at accretion time not
exceeding $2.6\%$ of the primary's final $M_{str}$. The satellite
marked with an arrow is a formal ``escaper'' for which $r_{\rm apo}$
cannot be computed.}
\label{fig:rhist}
\end{figure}
\end{center}

\section{Application to the Local Group}
\label{sec:LG}

We may apply these results to the interpretation of kinematical
outliers within the satellite population around the Milky Way (MW) and
M31, the giant spirals in the Local Group. Although part of the
discussion that follows is slightly speculative due to lack of
suitable data on the three-dimensional orbits of nearby satellites, we
feel that it is important to highlight the role that the concomitant
accretion of multiple satellites may have played in shaping the
dynamics of the dwarf members in the Local Group.

\subsection{Milky Way satellites}
\label{ssec:MW}

The filled squares in Figure~\ref{fig:rvMW} show the galactocentric
radial velocity of thirteen bright satellites around the Milky Way and
compare them with the simulated satellite population. This comparison
requires a choice for the virial radius and virial velocity of the
Milky Way, which are observationally poorly constrained.

We follow here the approach of Sales et al (2007), and use the
kinematics of the satellite population itself to set the parameters of
the Milky Way halo. These authors find that simulated satellites are
only mildly biased in velocity relative to the dominant dark matter
component: $\sigma_{\rm r} \sim 0.9 (\pm 0.2) V_{\rm vir}$, where
$\sigma_{\rm r}$ is the radial velocity dispersion of the satellite
population within $r_{vir}$. Using this, we find 
$V_{\rm vir}^{\rm MW}=109 \pm 22$ km/s
and $r_{\rm vir}^{\rm MW}=237\pm 50$ kpc from the observed radial
velocity dispersion of $\sim 98$ km/s. This corresponds to $M_{\rm
vir}^{\rm MW}=7 \times 10^{11} M_{\odot}$, in reasonable agreement
with the $1$-$2 \times 10^{12} M_{\odot}$ estimate of \citet{klypin02}
and with the recent findings of \citet{smith06} based on
estimates of the escape velocity in the solar neighbourhood.

Since Leo I dwarf has the largest radial velocity of the
Milky Way satellites, we have recomputed the radial velocity
dispersion excluding it from the sample. We have found that
$\sigma_r$ drops from 98 to 82 km/s when Leo I is not taken
into account changing our estimation of $V_{\rm vir}^{\rm MW}$ from 
109 to 91 km/s, still within the errors of the value previously found.
Given the recent rapid growth in the number of known Milky Way 
satellite one would suspect that the velocity dispersion 
will significantly increase if more Leo I-like 
satellites are detected. However, we notice that given their 
high velocities they are not expected to remain inside the
virial radius for a long time period hence not contributing 
to the $\sigma_r$ computation.

Figure~\ref{fig:rvMW} shows that, considering $V_{\rm vir}^{\rm MW}=109$ km/s,
the velocities and positions of all MW satellites are reasonably
consistent with the
simulated satellite population, with the possible exception of Leo I,
which is located near the virial radius and is moving outward with a
velocity clearly exceeding $V_{\rm vir}$. Indeed, for $V_{\rm
vir}^{\rm MW}=109$ km/s, Leo I lies right on the escape velocity curve
of an NFW profile with concentration parameter similar to those
measured in the simulations. This is clearly a kinematical outlier
reminiscent of the satellite expelled by three-body interactions
discussed in the previous subsection and identified by a dot-centered
circle in Figure~\ref{fig:rvMW}. This is the {\it only} ``associated''
satellite in our simulations with radial velocity exceeding $V_{\rm
vir}$ and located outside $r_{\rm vir}$.

Could Leo I be a satellite that has been propelled into a
highly-energetic orbit through a three-body interaction? If so, there
are a number of generic predictions that might be possible to verify
observationally. One is that its orbit must be now basically radial in
the rest frame of the Galaxy, although it might be some time before
proper motion studies are able to falsify this prediction.  A second
possibility is to try and identify the second member of the pair to
which it belonged. An outward moving satellite on a radial orbit takes
only $\sim 2$-$3$ Gyr to reach $r_{\rm vir}$ with escape
velocity. Coincidentally, this is about the time that the Magellanic
Clouds pair were last at pericenter, according to the traditional
orbital evolution of the Clouds \citep[see, e.g.,][]{gardiner96,vandermarel02}.

Could Leo I have been a Magellanic Cloud satellite ejected from the
Galaxy a few Gyrs ago? Since most satellites that are ejected do so
during {\it first} pericentric approach, this would imply that the
Clouds were accreted only recently into the Galaxy, so that they
reached their first pericentric approach just a few Gyr ago. This is
certainly in the spirit of the re-analysis of the orbit of the Clouds
presented recently by \citealt{besla07} and based on new proper
motion measurements recently reported by \citet{kallivayalil06}. In
this regard, the orbit of the Clouds might resemble the orbit of the
companion of the ``escaping'' satellite located next to Leo I in
Figure~\ref{fig:rvMW}. The companion is fairly massive and, despite a
turnaround radius of almost $\sim 600$ kpc and a rather late accretion
time ($t_{\rm acc}=10.5$ Gyr, see Figure~\ref{fig:orbesc}), it is left
after pericenter on a tightly bound, short-period orbit resembling
that of the Clouds today \citep{gardiner96,vandermarel02}. To
compound the resemblance, this satellite has, at accretion time, a
total luminosity of order $\sim 10\%$ of that of the primary, again on
a par with the Clouds.

We also note that an ejected satellite is likely to have picked up its
extra orbital energy through a rather close pericentric passage and
that this may have led to substantial tidal damage. This, indeed, has
been argued recently by \citealt{sohn06} on the basis of asymmetries
in the spatial and velocity distribution of Leo I giants (but see
\citealt{koch07} for a radically different interpretation).

On a final note, one should not forget to mention another (less
exciting!) explanation for Leo I: that our estimate of $V_{\rm
vir}^{\rm MW}$ is a substantial underestimate of the true virial
velocity of the Milky Way. The arrows in Figure~\ref{fig:rvMW}
indicate how the position of the MW satellites in this plane would
change if our estimate of $V_{\rm vir}^{\rm MW}$ is varied by $\pm
20\%$. Increasing $V_{\rm vir}^{\rm MW}$ by $\sim 20\%$ or more would
make Leo I's kinematics less extreme, and closer to what would be
expected for a high-speed satellite completing its first orbit. This
rather more prosaic scenario certainly cannot be discounted on the
basis of available data (see, e.g., Zaritsky et al 1989, Kochanek
1996, Wilkinson \& Evans 1999)

\begin{center}
\begin{figure}
\includegraphics[width=84mm]{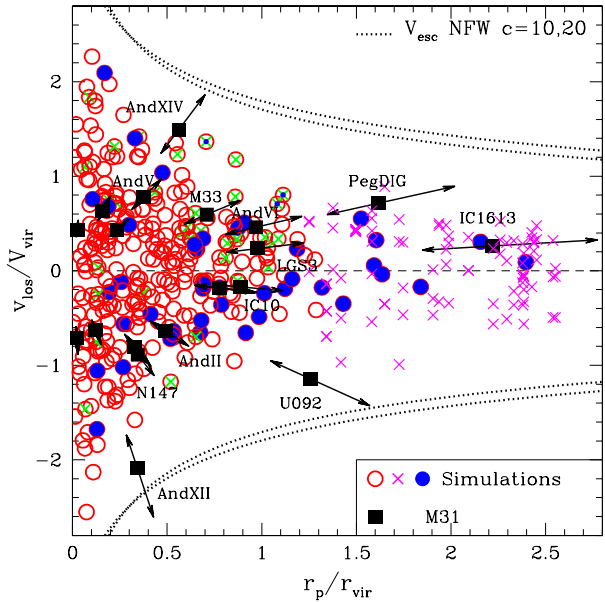}
\caption{As Figure~\ref{fig:rvMW} but for {\it line-of-sight}
velocities and {\it projected} distances. Three random orthogonal
projections have been chosen for each simulated satellite
system. Signs for $V_{\rm los}$ have been chosen so that it is
positive if the satellite is receding away from the primary in
projection, negative otherwise. The ``escaping'' satellite from
Figure~\ref{fig:rvMW} is shown by a starred symbol. Filled squares
correspond to the M31 satellites taken from \citealt{mcconnachieandirwin06},
plus And XIV \citep{majewski07} and And XII (Chapman et al 2007, submitted)
and assuming that $V_{\rm vir}^{\rm M31}\sim 138$ km/s and $r_{\rm
vir}^{\rm M31}=300$ kpc. Arrows indicate how the positions of M31
satellites in this plot would be altered if our estimate of $V_{\rm
vir}^{\rm M31}$ (and, consequently, $r_{\rm vir}^{\rm M31}$) is allowed
to vary by $20\%$.}
\label{fig:rvM31}
\end{figure}
\end{center}

\subsection{M31 satellites}
\label{ssec:M31}

A similar analysis may be applied to M31 by using the projected
distances and line-of-sight velocities of simulated satellites, shown
in Figure~\ref{fig:rvM31}. Three orthogonal projections of the
simulated satellites are overlapped in this figure, with symbols as
defined in Figure~\ref{fig:rvMW}. Following the same approach as in
\S~\ref{ssec:MW}, we use the fact that the line-of-sight satellite
velocity dispersion is $\sigma_{\rm los} \sim 0.8 (\pm 0.2) \, V_{\rm
vir}$ in our simulations to guide our choice of virial velocity and
radius for M31; $V_{\rm vir}^{\rm M31}=138 \pm 35$ km/s and $r_{\rm
vir}^{\rm M31}=300 \pm 76$ kpc. (We obtain $\sigma_{\rm los}=111$ km/s
for all $17$ satellites within $300$ kpc of M31.) This compares
favourably with the $V_{\rm vir}^{\rm M31}\sim 120$ km/s estimate
recently obtained by \citealt{seigar06} under rather different
assumptions.

With this choice, we show the $19$ satellites around M31 compiled by
McConnachie \& Irwin (2006), plus two recently-discovered satellites
for which positions and radial velocities have become available (And
XII, Chapman et al 2007, and And XIV, Majewski et al. 2007). 
As in Figure~\ref{fig:rvMW}, arrows
indicate how the position of M31 satellites would change in this
figure if $V_{\rm vir}^{\rm M31}$ were allowed to vary by $\pm
20\%$. We notice that the exclusion of And XII and And XIV (the
highest velocity satellites within 300 kpc from Andromeda) in the
$V_{\rm vir}^{\rm M31}$ estimation gives $\sim 100$ km/s, consistent
with the $V_{\rm vir}^{\rm M31}=138 \pm 35$ km/s previously found
considering all satellites.
Projected distances are as if viewed from infinity
along the direction joining the Milky Way with M31 and that the
{\it sign} of the line-of-sight velocity in Figure~\ref{fig:rvM31} is
chosen to be positive if the satellite is receding from the primary
(in projection) and negative otherwise.

There are a few possible outliers in the distribution of M31 satellite
velocities: And XIV (Majewski et al 2007), the Pegasus dwarf irregular
(UGC 12613, \citealt{gallagher98}), And XII (Chapman et al 2007), and UGCA 092
(labelled U092 in Figure~\ref{fig:rvM31}, \citealt{mcconnachieandirwin06}).
And XIV and PegDIG
seem likely candidates for the three-body ``ejection'' mechanism
discussed above: they have large velocities for their position, and,
most importantly, they are receding from M31; a {\it requirement} for
an escaping satellite. Note, for example, that And XIV lies very close
to the ``escaping'' satellite (dot-centered symbol in
Figure~\ref{fig:rvM31}) paired to Leo I in the previous
subsection. Escapers should move radially away from the primary, and
they would be much harder to detect in projection as extreme velocity
objects, unless they are moving preferentially along the line of
sight. It is difficult to make this statement more conclusive without
further knowledge of the orbital paths of these satellites.  Here, we
just note, in agreement with Majewski et al (2007), that whether And
XIV and PegDIG are dynamical ``rogues'' depends not only on the
(unknown) transverse velocity of these galaxies, but also on what is
assumed for M31's virial velocity. With our assumed $V_{\rm vir}^{\rm
M31}=138$ km/s, neither And XIV nor PegDIG look completely out of
place in Figure~\ref{fig:rvM31}; had we assumed the lower value of
$120$ km/s advocated by Seigar et al (2006) And XIV would be almost
on the NFW escape velocity curve, and would certainly be a true
outlier.

High-velocity satellites {\it approaching} M31 in projection are
unlikely to be escapers, but rather satellites on their first
approach.  This interpretation is probably the most appropriate for
And XII and UGCA 092. As discussed by Chapman et al (2007), And XII is
almost certainly {\it farther} than M31 but is approaching us at much
higher speed ($\sim 281$ km/s faster) than M31. This implies that And
XII is actually getting closer in projection to M31 (hence the
negative sign assigned to its $V_{\rm los}$ in
Figure~\ref{fig:rvM31}), making the interpretation of this satellite
as an escaping system rather unlikely.

Note, again, that although And XII (and UGCA 092) are just
outside the loci delineated by simulated satellites in
Figure~\ref{fig:rvM31}, revising our assumption for $V_{\rm vir}^{\rm
M31}$ upward by $20\%$ or more would render the velocity of this
satellite rather less extreme, and would make it consistent with that
of a satellite on its first approach to M31. As was the case for Leo
I, this more prosaic interpretation of the data is certainly
consistent with available data.

\section{SUMMARY and Conclusions}
\label{sec:conc}

We examine the orbits of satellite galaxies in a series of
Nbody/gasdynamical simulations of the formation of $L_*$ galaxies in a
$\Lambda$CDM universe. Most satellites follow orbits roughly in accord
with the expectations of secondary infall-motivated models. Satellites
initially follow the universal expansion before being decelerated by
the gravitational pull of the main galaxy, turning around and
accreting onto the main galaxy. Their apocentric radii decrease
steadily afterwards as a result of the mixing associated with the
virialization process as well as of dynamical friction. At $z=0$ most
satellites associated with the primary are found within its virial
radius, and show little spatial or kinematic bias relative to the dark
matter component (see also Sales et al 2007).

A number of satellites, however, are on rather unorthodox orbits, with
present apocentric radii exceeding their turnaround radii, at times
by a large factor. The apocenters of these satellites are typically
beyond the virial radius of the primary; one satellite is formally
``unbound'', whereas another is on an extreme orbit and is found today
more than $2.5\, r_{\rm vir}$ away, or $\gsim \, 600$ Mpc when scaling
this result to the Milky Way.

These satellites owe their extreme orbits to three-body interactions
during first approach: they are typically the lighter member of a pair
of satellites that is disrupted during their first encounter with the
primary. This process has affected a significant fraction of
satellites: a full one-third of the simulated satellite population
identified at $z=0$ have apocentric radii exceeding their turnaround
radii. These satellites make up the majority ($63\%$) of systems on
orbits that venture outside the virial radius.

We speculate that some of the kinematical outliers in the Local Group
may have been affected by such process. In particular, Leo I might
have been ejected $2$-$3$ Gyr ago, perhaps as a result of interactions
with the Milky Way and the Magellanic Clouds. Other satellites on
extreme orbits in the Local Group may have originated from such
mechanism. Cetus \citep{lewis07} and Tucana \citep{oosterloo96}
---two dwarf spheroidals in the periphery
of the Local Group---may owe their odd location (most dSphs are found
much closer to either M31 or the Galaxy) to such ejection mechanism.

If this is correct, the most obvious culprits for such ejection events
are likely to be the largest satellites in the Local Group (M33 and
the LMC/SMC), implying that their possible role in shaping the
kinematics of the Local Group satellite population should be
recognized and properly assessed.  In this regard, the presence of
kinematical oddities in the population of M31 satellites, such as the
fact that the majority of them lie on ``one side'' of M31 and seem to
be receding away from it (McConnachie \& Irwin 2006), suggest the
possibility that at least some of the satellites normally associated
with M31 might have actually been brought into the Local Group fairly
recently by M33. Note, for example, that two of the dynamical outliers
singled out in our discussion above (And XII and And XIV) are close to
each other in projection; have rather similar line-of-sight velocities
(in the heliocentric frame And XII is approaching us at $556$ km/s, And
XIV at $478$ km/s); and belong to a small subsystem of satellites
located fairly close to M33.

The same mechanism might explain why the spatial distribution of at
least some satellites, both around M31 and the Milky Way, seem to
align themselves on a ``planar'' configuration \citep{majewski94,libeskind05,
kochandgrebel06}, as this may
just reflect the orbital accretion plane of a multiple system of
satellites accreted simultaneously in the recent past \citep{kroupa05,
metz07}.

From the point of view of hierarchical galaxy formation models, it
would be rather unlikely for a galaxy as bright as M33 to form in
isolation and to accrete as a single entity onto M31. Therefore, the
task of finding out {\it which} satellites (rather than {\it whether}) have
been contributed by the lesser members of the Local Group, as well as
what dynamical consequences this may entail, should be undertaken
seriously, especially now, as new surveys begin to bridge our incomplete
knowledge of the faint satellites orbiting our own backyard.

\section*{Acknowledgements}
\label{acknowledgements}

LVS and MGA are grateful for the hospitality of the Max-Planck
Institute for Astrophysics in Garching, Germany, where much of the
work reported here was carried out. LVS thanks financial support from
the Exchange of Astronomers Programme of the IAU and to the ALFA-LENAC
network. JFN acknowledges support from Canada's NSERC, from the
Leverhulme Trust, and from the Alexander von Humboldt Foundation, as
well as useful discussions with Simon White, Alan McConnachie, and
Jorge Pe\~narrubia. MS acknowledges support by the German Science
foundation (DFG) under Grant STE 710/4-1. We thank Scott Chapman and
collaborators for sharing their results on Andromeda XII in advance of
publication. We also acknowledge a very useful report from an anonymous 
referee that helped to improve the first version. 

\bibliography{references}

\end{document}